\documentclass[conference]{IEEEtran}
\IEEEoverridecommandlockouts
\usepackage{cite}
\usepackage{graphicx}
\usepackage{textcomp}
\usepackage{eso-pic} 
\def\BibTeX{{\rm B\kern-.05em{\sc i\kern-.025em b}\kern-.08em
    T\kern-.1667em\lower.7ex\hbox{E}\kern-.125emX}}
\begin{document}
\AddToShipoutPictureFG*{%
  \AtPageLowerLeft{%
    \put(\LenToUnit{0.5\paperwidth},\LenToUnit{0.32in}){%
      \makebox[0pt][c]{\parbox{\textwidth}{\centering\scriptsize
        \textcopyright~2026 IEEE. Personal use of this material is permitted. Permission from IEEE must be obtained for all other uses, in any current or future media, including reprinting/republishing this material for advertising or promotional purposes, creating new collective works, for resale or redistribution to servers or lists, or reuse of any copyrighted component of this work in other works.\\
        Accepted to the 2026 IEEE Frontiers in Education Conference (FIE). DOI: TBD}}}}}
\title{WIP: DBWorkout: A Gamified SQL Practice Platform to Support Formative Learning in Database Courses}
\author{\IEEEauthorblockN{Sehrish Basir Nizamani, Deepika Devaraj, Tien Nguyen, Khyati Goyal, \\
Saad Nizamani, Sally Hamouda, and Jaren Goldberg}
\IEEEauthorblockA{\textit{Department of Computer Science} \\
\textit{Virginia Tech}\\
Blacksburg, VA, USA \\
\{sehrishbasir, deepika, ntien22, khyati22, saadnizamani, sallyh84, jaren\}@vt.edu}
}
\maketitle
\begin{abstract}
This research WIP paper presents DBWorkout, a web-based platform that supports formative SQL learning through sandbox-based execution, automated result-based feedback, and session-based gamification. Learning Structured Query Language (SQL) remains challenging for undergraduate students due to limited opportunities for interactive practice and immediate feedback. Students iteratively practice SQL on live database instances while receiving multi-dimensional feedback on query correctness, including row values, column structure, and ordering. To reduce instructor workload, DBWorkout incorporates large language model (LLM)-assisted tools for schema and task generation within a human-in-the-loop workflow. A pilot study with teaching assistants and a classroom deployment involving 170 undergraduate students across two in-class sessions show strong perceived learning value (90\% agreement) and engagement (87\% enjoyment), alongside low reported pressure (22\%). However, only 42\% of students found the automated feedback sufficiently actionable, a finding independently corroborated by 40\% of open-ended responses raising feedback quality concerns, providing cross-method triangulation of this gap. These findings demonstrate the technical feasibility and early pedagogical potential of DBWorkout while identifying directions for enhancing feedback quality and supporting sustained SQL learning.
\end{abstract}


\section{Introduction}
This Work-in-Progress (WIP) research paper presents DBWorkout, a computer-based instructional platform designed to support formative Structured Query Language (SQL) practice and student engagement in undergraduate database courses. Learning SQL remains challenging for many students due to limited opportunities for iterative practice and the lack of immediate, meaningful feedback in traditional instructional tools. Students often struggle with core concepts such as JOIN operations and GROUP BY clauses, leading to difficulties in translating conceptual understanding into correct query formulation \cite{yang2025sql}. Maintaining engagement during SQL practice is also difficult because traditional approaches rely on static assignments and isolated practice with limited interactivity, feedback, or motivational support. Providing feedback that is both accurate and actionable therefore remains an open challenge, one our preliminary findings identify as a key area for improvement in DBWorkout.
To address these challenges, we introduce DBWorkout, a web-based platform that supports iterative SQL practice through a gamified, sandbox-oriented environment. The platform provides automated, result-based feedback within an isolated execution environment, allowing students to experiment and refine queries incrementally. Gamified features, including real-time leaderboards, are designed to promote engagement and encourage active participation during in-class practice. In addition, DBWorkout incorporates large language model (LLM) support to assist instructors in generating database schemas, practice questions, and reference SQL solutions, reducing content creation overhead.
This WIP paper is guided by three research questions:
\begin{itemize}
    \item \textbf{RQ1:} How do students perceive the usability and learning value of DBWorkout during in-class SQL practice?
    \item \textbf{RQ2:} Does the session-bounded leaderboard support engagement without inducing performance anxiety?
    \item \textbf{RQ3:} What aspects of automated feedback do students find insufficient, and what improvements are indicated?
\end{itemize}
This paper presents the design, implementation, and preliminary empirical evaluation of DBWorkout. Following an internal pilot with teaching assistants, the platform was deployed in an introductory database management course with approximately 170 undergraduate students across two in-class sessions. We report preliminary findings from a post-session survey examining student engagement, perceived learning value, and user experience, and identify key directions for future development.
\section{Related Work and Instructional Challenges}
Research on SQL learning tools spans several decades, yet challenges around iterative practice, immediate feedback, and instructor adaptability remain unresolved. Wen et al. \cite{wen2024sqlplatform} developed a browser-based SQL platform with automated grading, and noted that other browser-based platforms such as SQLbolt and Programiz suffer from fixed database structures. Mitrovic and Ohlsson \cite{mitrovic2015sqltutor} demonstrated the effectiveness of SQL-Tutor, though constraint-based tutors require substantial upfront effort to encode a new domain's syntax and semantic constraints before any content can be authored.
Gamification has been proposed as a means of improving student engagement in computing courses, though its effectiveness is context-dependent. Ahmad et al. \cite{ahmad2020gamification} found across 229 undergraduate computer science students that gamified sections statistically significantly outperformed non-gamified control sections,  with improvements over the semester sustained for individual and small-group students, though less consistent for large groups. Li et al. \cite{li2024leaderboards}, in a systematic review synthesizing 20 articles (22 studies; 29 interventions) in higher education, found that motivational effects were not uniform: all interventions reporting positive motivational effects used absolute leaderboards, and separately, providing transparent scoring rules was also associated with positive motivation and performance effects. Critically, whether leaderboards that persist across a semester affect lower-ranked students' motivation over time remains empirically untested, Li et al. \cite{li2024leaderboards} note that longitudinal and novelty effects of leaderboard use have not been extensively examined in the literature. As a precaution, DBWorkout implements a session-bounded absolute leaderboard that resets each class session, with transparent point-awarding based on query correctness. This design is specifically motivated by the need to sustain participation in each in-class practice session without accumulating discouragement over a semester-long ranking history.
Keuning et al. \cite{keuning2019feedback}, in a systematic review of 101 automated feedback tools for programming exercises, found that nearly all tools focus on identifying mistakes (96\%), while fewer than 20\% provide explicit task-processing-step guidance, and that teachers cannot easily adapt most tools to their own needs. DBWorkout goes beyond correctness reporting by providing result-based feedback that surfaces specific error patterns, including those associated with JOIN and aggregation queries. Prior work has raised prompt sensitivity and hallucination as general risks of unguided LLM use \cite{prakash2024llmdb}. DBWorkout addresses both for instructor content generation through a human-in-the-loop workflow for content creation and validation.
SQL presents instructional challenges that distinguish it from other introductory programming topics. Mitrovic and Ohlsson \cite{mitrovic2015sqltutor} noted that SQL queries require understanding an underlying database schema, interpreting ambiguous natural language problem descriptions, and navigating a language where multiple syntactically valid solutions exist. The most consistently documented challenge is student difficulty with JOIN and GROUP BY operations. Alkhabaz et al. \cite{alkhabaz2023query} analyzed over 357,000 query submissions from 462 students and found that every concept labeled as GroupBy produced above-average syntax error counts, and that advanced JOIN and subquery problems generated the highest combined error totals of any problem category. This evidence motivates two specific design choices in DBWorkout: (1) feedback messages that explicitly surface JOIN- and aggregation-related error patterns rather than reporting only generic row count mismatches, and (2) operation-type tagging of every practice task so instructors can target specific problem categories. DBWorkout addresses the broader gap by integrating sandbox execution, actionable feedback, gamified engagement, and low-friction authoring within a single platform.
\section{System Design}
\subsection{Design Goals}
\begin{figure*}[t]
\centering
\includegraphics[width=0.8\textwidth]{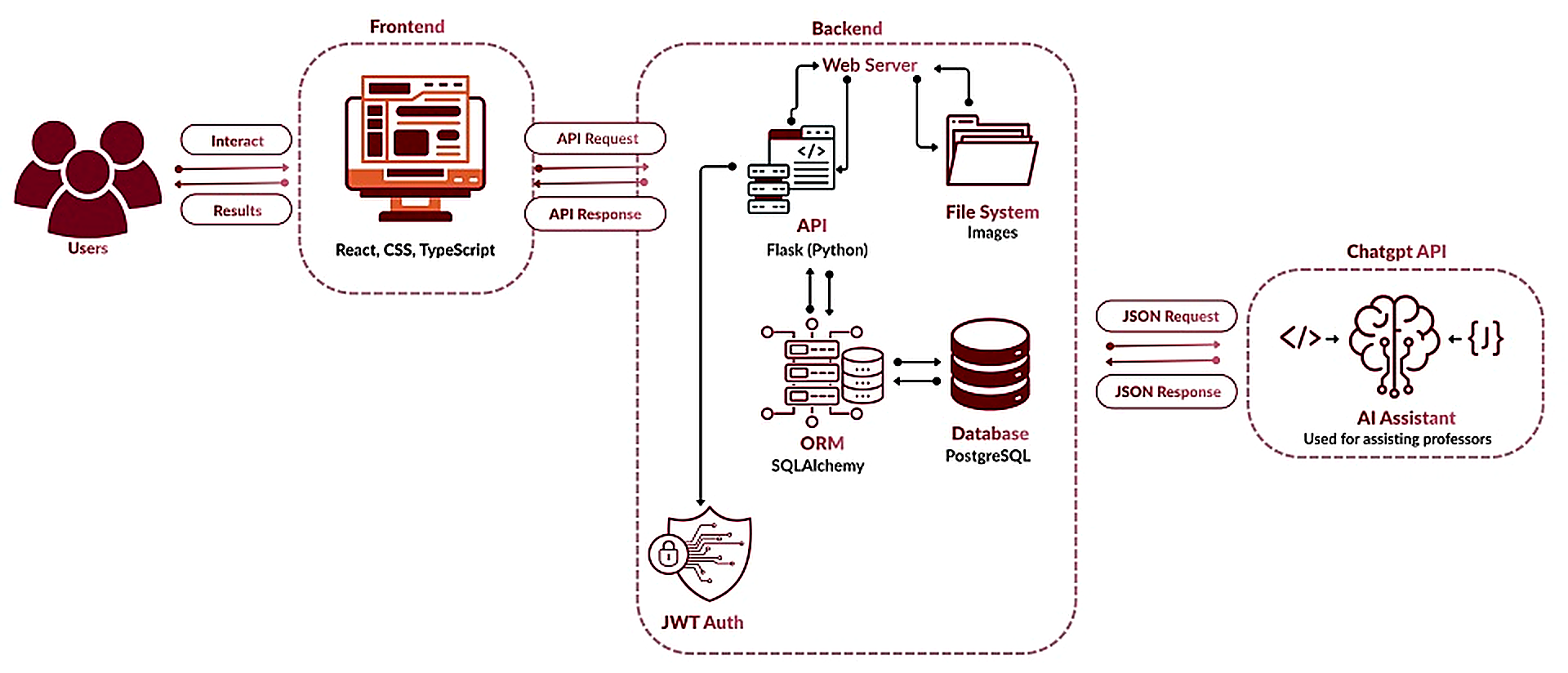}
\caption{DBWorkout system architecture and sandbox execution flow.}
\label{fig:dbworkout}
\end{figure*}

DBWorkout targets five barriers to SQL instruction: authentic iterative practice on a live database, multi-dimensional feedback supporting partial credit and refinement, controlled sandbox execution isolating each submission, low-friction instructor authoring without manual grading logic, and session-bounded gamification to motivate participation without persistent ranking pressure.
DBWorkout follows a three-tier architecture. The frontend is a React single-page application (TypeScript, Vite) that communicates with a Flask REST API backend over HTTPS. The backend connects to a PostgreSQL 16 database that serves dual roles: storing application data (users, courses, tasks, and submissions) and hosting student sandbox schemas for SQL execution. Authentication uses JSON Web Tokens (JWT) with a sliding refresh window. An integrated Monaco Editor provides syntax highlighting, autocompletion, and inline error display.
The backend checks for the request's JWT token at every endpoint to ensure that users can only access authorized resources. All user inputs are sanitized and validated before execution. Rather than provisioning separate database instances per student, DBWorkout exploits PostgreSQL schema namespacing to create lightweight isolated environments. Each submission executes in a cloned sandbox with restricted search paths, per-statement timeouts, and controlled execution limits, after which the sandbox is dropped. This approach achieves sub-second sandbox creation while providing strong isolation within a single PostgreSQL instance, as illustrated in Fig.~\ref{fig:dbworkout}.
\subsection{Automated Feedback Mechanism}
DBWorkout employs result-based feedback: the system executes the instructor's reference SQL and the student's submission in identical sandbox environments, then compares results across seven weighted dimensions --- execution success, statement count, column names, column types, row count, row values (multiset equality), and ordering (when applicable). Values are normalized before comparison; each check produces a descriptive feedback string such as ``Missing 3 rows'' or ``Column \texttt{salary} has type VARCHAR, expected numeric,'' supporting iterative refinement. Optional method-requirement checks (e.g., requiring a JOIN) can be enforced as pass/fail constraints. For Data Manipulation Language (DML) tasks, the system compares table-state snapshots before and after execution; for Data Definition Language (DDL) tasks, structural comparison verifies column definitions and constraints. The reference result is cached and automatically recomputed when the underlying schema changes.
\subsection{Gamification and Leaderboard Integration}
DBWorkout integrates a time-competitive leaderboard into instructor-managed practice sessions. The live leaderboard ranks students by number of tasks completed (descending) and cumulative time to completion (ascending), measured from session start to each task's first correct submission. The frontend displays animated rank changes and a podium visualization for the top three students. Confetti animations fire when a session concludes, and the leaderboard auto-refreshes via polling to reflect new submissions in near real-time. There are no persistent points, badges, or streaks; the motivation is per-session and task-focused. This reflects the same precautionary design rationale discussed in Section II: resetting rankings each session ensures every student begins on equal footing, regardless of whether accumulated low standing would prove demotivating.
\subsection{LLM-Supported Instructor Authoring Tools}
Driven by the need to automate time-intensive instructional tasks and minimize the high costs of traditional course administration \cite{prakash2024llmdb}, DBWorkout integrates GPT-4o for two instructor-facing workflows, both gated behind instructor-level authorization. For task generation, instructors specify a schema, difficulty level, and free-text prompt; the model returns 5--10 candidate tasks (title, description, reference SQL), which the instructor reviews and approves before deployment. For schema creation, instructors describe a domain in natural language and receive a complete schema with data; additional rows maintaining referential integrity can be generated on demand. All content is validated against the schema and approved by the instructor prior to student use, maintaining pedagogical control while mitigating hallucination risks identified in prior work \cite{prakash2024llmdb}.
\section{Pilot Study}
Seven Teaching Assistants (TAs) evaluated platform readiness prior to classroom deployment, with 2 acting as instructors and 5 as students following a scripted protocol. In the instructor role, TAs performed end-to-end course management including schema generation and LLM-assisted task curation. In the student role, TAs completed SQL tasks under simulated classroom conditions. TAs subsequently submitted written defect and usability reports, which the research team categorized to prioritize refinements. Three defects were identified and resolved (reference-solution validation, feedback persistence on navigation, enrollment timeouts), after which the sandbox, leaderboard, and LLM tools performed as intended.
\section{Classroom Deployment and Methods}
\subsection{Study Context}
Following the pilot, DBWorkout was deployed in an introductory undergraduate database management course at a large research university. Two in-class practice sessions were conducted: a morning session ($n=104$) and an evening session ($n=66$), totaling 170 students. In each session, the instructor introduced DBWorkout, students logged in using pre-provisioned accounts, and a timed practice session began. Students worked individually to complete SQL tasks in the leaderboard-enabled environment. The tasks were authored by the course instructor, optionally assisted by the LLM authoring tool, and reviewed by a faculty member prior to deployment.
\subsection{Survey Instruments}
To address RQ1--RQ3, a post-activity survey was administered immediately after each session. The survey comprised items adapted from two validated instruments and a set of system-specific items. The User Experience Questionnaire (UEQ) \cite{laugwitz2008ueq} provided distinct subscales evaluating pragmatic attributes alongside creative factors like novelty; its semantic differential format is well-suited to capturing first-impression system evaluations. The Intrinsic Motivation Inventory (IMI) \cite{ryan1982cognitive} subscales for enjoyment, perceived competence, effort, perceived value, and pressure/tension were included because IMI has been widely used to assess intrinsic motivation in educational technology contexts and its subscales directly map to the engagement and anxiety dimensions relevant to RQ2. Likert-scale responses were coded 1--5. An open-ended item asked students to describe their experience and suggest improvements.
\subsection{Qualitative Analysis}
Open-ended responses ($n=170$; 98 retained as substantive after removing blanks, single-word answers, and off-topic notes) were analyzed using reflexive thematic analysis \cite{braun2019reflecting}. One researcher read all responses and developed an initial coding scheme; codes were grouped into candidate themes. A second researcher independently reviewed and coded the data, with disagreements resolved through discussion \cite{thomas2006inductive}. The credibility of identified themes is further supported by convergence with the quantitative findings: the dominant theme of feedback depth and specificity arose independently across 40\% of open-ended responses, closely aligning with the quantitative finding that only 42\% found automated feedback sufficient.
\section{Preliminary Findings}
\subsection{Quantitative Findings}
Table~\ref{tab:survey} summarizes descriptive statistics for all survey items ($n=170$). For UEQ items, \% Positive indicates ratings of 4 or 5 on the bipolar scale. For IMI and system-specific items, \% Positive indicates Agree or Strongly Agree.
\begin{table}[t]
\centering
\caption{Survey Item Descriptive Statistics ($n=170$)}
\label{tab:survey}
\begin{tabular}{|l|c|c|c|}
\hline
\textbf{Item} & \textbf{Mean} & \textbf{SD} & \textbf{\% Pos.} \\
\hline
\multicolumn{4}{|l|}{\textit{UEQ -- Usability}} \\
\hline
Bad--Good & 4.42 & 0.60 & 95\% \\
Unappealing--Appealing & 4.43 & 0.66 & 94\% \\
Not understandable--Understandable & 4.27 & 0.77 & 86\% \\
Complicated--Easy & 3.69 & 0.97 & 60\% \\
Confusing--Clear & 4.01 & 0.90 & 77\% \\
Slow--Fast & 4.16 & 0.89 & 79\% \\
\hline
\multicolumn{4}{|l|}{\textit{UEQ -- Novelty}} \\
\hline
Usual--Leading edge & 3.71 & 0.84 & 61\% \\
Conventional--Inventive & 3.78 & 0.91 & 62\% \\
\hline
\multicolumn{4}{|l|}{\textit{IMI -- Enjoyment}} \\
\hline
I enjoyed using DBWorkout & 4.13 & 0.65 & 87\% \\
This activity was fun to do & 4.12 & 0.61 & 87\% \\
I would use this again & 4.09 & 0.79 & 83\% \\
\hline
\multicolumn{4}{|l|}{\textit{IMI -- Perceived Competence}} \\
\hline
I felt capable while using DBWorkout & 3.91 & 0.75 & 76\% \\
I think I performed well & 3.74 & 0.74 & 65\% \\
\hline
\multicolumn{4}{|l|}{\textit{IMI -- Effort}} \\
\hline
I put a lot of effort into this & 3.94 & 0.74 & 73\% \\
I tried hard while using DBWorkout & 3.91 & 0.76 & 73\% \\
\hline
\multicolumn{4}{|l|}{\textit{IMI -- Value}} \\
\hline
Useful for learning SQL & 4.32 & 0.71 & 90\% \\
Helps understand course concepts & 4.21 & 0.65 & 87\% \\
Value in using for future learning & 4.30 & 0.61 & 92\% \\
\hline
\multicolumn{4}{|l|}{\textit{IMI -- Pressure/Tension}} \\
\hline
I felt pressured & 2.75 & 0.92 & 22\% \\
I felt frustrated & 2.98 & 0.98 & 35\% \\
\hline
\multicolumn{4}{|l|}{\textit{System-Specific}} \\
\hline
Feedback sufficient to identify mistakes & 3.18 & 1.09 & 42\% \\
Schema details sufficient & 3.85 & 0.72 & 76\% \\
\hline
\end{tabular}
\end{table}
\textit{RQ1 -- Usability and learning value.} Students rated DBWorkout positively across all usability dimensions (UEQ subscale $M=4.16$). The weakest item was Easy over Complicated ($M=3.69$; 60\% positive), suggesting that some students experienced complexity in the tasks or platform workflow. Novelty scores were moderate ($M=3.71$--$3.78$), indicating that DBWorkout was perceived as contextually familiar rather than highly novel. Perceived value items produced the highest scores ($M=4.21$--$4.32$), with 90--92\% agreement; 83\% indicated they would use DBWorkout again.
\textit{RQ2 -- Engagement and pressure.} Enjoyment was strong (87\%), and pressure was low ($M=2.75$; 22\%), supporting the interpretation that the session-bounded leaderboard preserved motivational benefits without generating performance anxiety.
\textit{RQ3 -- Feedback sufficiency.} The lowest-scoring item was feedback sufficiency ($M=3.18$, $SD=1.09$), with only 42\% finding feedback adequate. Frustration was moderate ($M=2.98$; 35\%). Schema quality was rated more positively ($M=3.85$; 76\%).
\subsection{Qualitative Findings}
Six themes emerged from reflexive thematic analysis of the 98 substantive open-ended responses.
\textit{Feedback depth and specificity (40\% of responses):} Error messages lacked sufficient direction. ``I did not feel any more knowledgeable going into my second or third attempts because the feedback was vague and uninformative.''
\textit{Positive reception (17\%):} Students expressed enjoyment and compared DBWorkout favorably to existing platforms. ``Enjoyed using the app -- felt sort of like LeetCode for SQL.''
\textit{UX and interface issues (16\%):} Missing auto-save, schema navigation challenges, and occasional lag.
\textit{Task clarity (12\%):} Ambiguous wording and unclear expectations.
\textit{Attempt limit design (6\%):} Preference for continued access without penalties after limits.
\textit{Competitive features (9\%):} Mixed reactions to leaderboard; some found it motivating, others preferred collaborative or non-ranked practice.
\section{Discussion}
\subsection{RQ1: Usability and Perceived Learning Value}
Perceived value items produced the highest scores in the survey, providing initial empirical support for DBWorkout's formative design. The feedback sufficiency gap reported above is consistent with Keuning et al.'s finding that automated feedback tools rarely progress beyond identifying mistakes to guiding next steps \cite{keuning2019feedback}. Because this gap emerged independently from a structured Likert item and from unprompted open-ended responses, it is likely a robust finding rather than an artifact of survey design. Enhancing feedback specificity -- for example by displaying query output or providing directional hints for common errors such as missing JOIN conditions -- is the most clearly indicated next step.
\subsection{RQ2: Gamification and Engagement}
The high enjoyment and low pressure reported above confirm that the session-bounded leaderboard preserved motivational benefits without generating performance anxiety, consistent with Li et al.'s finding that absolute leaderboards are associated with positive motivational effects \cite{li2024leaderboards}. The qualitative data, however, revealed a minority of students who preferred non-competitive or collaborative formats, motivating future investigation of alternative engagement mechanics such as team-based challenges.
\subsection{RQ3: Feedback Actionability and Design Implications}
The feedback-actionability gap identified above is the most significant finding from this deployment. In the context of Keuning et al. \cite{keuning2019feedback}, this positions DBWorkout among the majority of tools that successfully identify mistakes but fall short of providing next-step guidance. Several enhancements targeting this gap have already been implemented and are described in the Future Work section.
\subsection{Scalability}
The sandbox architecture proved stable across 170 concurrent students with no infrastructure failures. The two-session deployment demonstrates that the platform scales to authentic classroom conditions, addressing a practical concern for adoption at course scale.
\section{Limitations}
The data reflect a single course at one institution following students' first exposure to DBWorkout; findings may not generalize to other contexts, and engagement levels may change with sustained use. The study measures perceived learning value rather than objective learning gains; pre- and post-assessments were not administered, and whether DBWorkout produces measurable improvements in SQL proficiency remains to be established. Finally, the qualitative analysis was conducted by members of the research team, introducing potential researcher bias, which was partially mitigated through independent second-coder review.
\section{Future Work}
The next development cycle targets two areas. \textit{Feedback enhancement:} Implemented improvements include value-level cell diffs, directional hints for common errors (Cartesian products, missing GROUP BY, NULL equality), jargon tooltips, and score-based closeness messages. These will be evaluated in the next deployment via pre- and post-assessments alongside feedback sufficiency ratings and attempt-to-correct rates. \textit{Inclusive engagement:} We will investigate team-based pair competition and disability-aware time normalization so leaderboard rankings do not disadvantage students with approved accommodations.
\section{Conclusion}
This paper presented DBWorkout, a gamified, sandbox-based SQL practice platform designed to support formative learning in undergraduate database courses. A pilot study and classroom deployment with 170 students demonstrated strong perceived learning value, high engagement, and low performance anxiety. The primary gap identified is feedback actionability: only 42\% of students found the automated feedback sufficient, a finding robustly supported by independent qualitative evidence. The session-bounded leaderboard design successfully balanced motivational benefits against the demotivation risks associated with persistent ranking systems. Future work will evaluate enhanced feedback mechanisms and inclusive competition features in a controlled deployment with pre- and post-assessment measures. DBWorkout represents a step toward integrating authentic practice, scalable feedback, and principled engagement mechanics within a single platform for SQL education.
\section*{Acknowledgment}
This project was supported in part by a TLOS TEL grant at Virginia Tech. The authors used LLM assistance for grammar correction and improving sentence structure in the preparation of this manuscript.
\bibliographystyle{IEEEtran}
\bibliography{references}
\end{document}